 \documentclass{ws-ijmpa}

\usepackage[super,compress]{cite}
\usepackage{graphicx}
\usepackage{hyperref}
\begin{document}
\markboth{Dariusz G\'ora for the Pierre Auger Collaboration}{}

%
\catchline{}{}{}{}{}
%

\title{Particle physics in cosmic rays
}

\author{Dariusz G\'ora for the Pierre Auger Collaboration~\footnote{Full author list:
\href{http://www.auger.org/archive/authors_2019_09.html}{http://www.auger.org/archive/authors\_2020\_09.html}}
\footnote{Observatorio Pierre Auger, Av. San Martín Norte 304, 5613 Malargue}  \footnote{\href{mailto:auger_spokespersons@fnal.gov}{\rm auger\_spokespersons@fnal.gov} }
}

\address{ Institute of Nuclear Physics Polish Academy of Sciences\\
 Radzikowskiego 152, 31-342 Kraków, Poland\\
Dariusz.Gora@ifj.edu.pl}

\maketitle

\begin{history}
\received{Day Month Year}
\revised{Day Month Year}
\end{history}

\begin{abstract}

The Pierre Auger Observatory is the world's largest detector of ultra--high energy cosmic rays (UHECRs). It uses an array of fluorescence telescopes and particle detectors at the ground to obtain detailed measurements of the energy spectrum, mass composition and arrival directions of primary cosmic rays (above the energy of $10^{17}$ eV) with accuracy not attainable until now.

Observations of extensive air showers performed by the Pierre Auger Observatory can also be used to probe hadronic interactions at high energy, in a kinematic and energy region not accessible by man-made accelerators. Indeed, exploiting Auger data, we reach center-of-mass energies up to 400 TeV, i.e. more than 30 times of those attainable at the LHC, and explore interactions in the very forward region of phase space on targets of atomic mass of 14. In addition, a precise measurement of the muon component of air showers at the ground is more sensitive to the details of the hadronic interactions along many steps of the cascade development, such as the multiplicity of the secondaries and the fraction of electromagnetic component with respect to the total signal. On the other hand, the intrinsic muon fluctuations mostly depend on the first interaction.

In this paper we overview the new Auger studies exploring the connection between the dynamics of the air shower and the multi-particle production, and how this knowledge can be translated into constraints of the high energy hadronic models as well as direct measurements, complementary to, and beyond the reach of, accelerator experiments.

\end{abstract}

\ccode{PACS numbers:}


\section{Introduction}

The discovery of cosmic rays by Victor Hess in 1912 opened a new window on the Universe\cite{hess} , providing a new channel of observation for some of the most energetic events occurring in the cosmos. In the twentieth century, many experiments were conducted to detect cosmic rays in an effort to understand their origin and nature. The data collected so far  by the leading cosmic ray experiments, like the Pierre Auger Observatory (Auger) or  Telescope Array~\cite{ta} , have greatly advanced our understanding of ultra-high energy cosmic rays (i.e. those with energies above 1 EeV = $10^{18}$ eV), but also revealed new, more specific questions. As an example  the suppression of the cosmic ray flux around 40 EeV has recently  been  established unambiguously~\cite{auger_spectrum} .
Such a suppression was predicted more than 50 years ago as a result of particle energy losses en route to Earth due to the pion photo-production on the cosmic microwave background and is known as the Greisen-Zatsepin-Kuzmin (GZK) cutoff~\cite{greisen,kuzmin} . However, data on fluctuations in the development of extensive air showers as well as other composition-sensitive observables require consideration of a different interpretation of the Auger data: that the observed flux suppression may be a manifestation of an upper-limit of the efficiency of the cosmic ray accelerators, rather than the GZK effect. In this case, the composition determination of UHECRs (preferably on event-by-event basis) is the key to resolve this issue.

Above $10^{14}$ eV, studies of  cosmic rays are possible  through observations of extensive air showers (EAS) induced by the cosmic rays. The first interactions of primary cosmic
rays take place high in the atmosphere, therefore we are not able to observe them directly. What can be observed are the cascades of secondary particles
moving in the atmosphere and finally reaching the ground. When ultra-high energy particles (protons, nuclei, photons or neutrinos) interact with the oxygen and the nitrogen molecules composing the atmosphere, they initiate cascades of secondary particles whose characteristics are a function of the properties of the atmospheric layer they propagate through, and of the type of physical interactions occurring at different stages of the cascade development. In the initial phase of the cascading process, the number of particles
increases while the energy per particle drops and distinct components emerge, namely the hadronic, electromagnetic and muonic components. Such growth carries on until a maximum is reached, as particles below a certain energy threshold are no longer capable of producing additional particles and as atmospheric absorption processes, such as ionization, take over. As many as $10^6-10^9$ secondary particles may reach the ground over an area that can extend up to several square kilometers. 

As we can see from the above  description, the properties of the primary particle must be deduced from the properties of the shower. Therefore, modeling of shower development in the atmosphere is an essential part of any UHECRs studies. However, the energy range relevant to the ultra-high energy cosmic ray studies is far beyond the energies accessible at terrestrial particle accelerators, at which the properties of hadronic interactions have been studied~\footnote{Exploiting UHECRs data, we reach center-of-mas  energies up to 400 TeV i.e. more than 30 times of those attainable at the  Large Hadron Collider (LHC).}. Thus, an extrapolation of hadronic interaction properties to higher energies is necessary, contributing to systematic uncertainties of the final results. In general,  extensive air shower simulations have to be made and compared to the experimental distributions. Clearly, the results depend critically on the accuracy of these simulations, which in turn depend on the assumptions made concerning interaction cross sections, inelasticity of interactions, multiplicities and distributions of produced particles, etc. Moreover, the present extrapolation of hadronic interaction properties to higher energies is done based on  
 a different phase space measurement in    accelerator experiments, than required for air showers i.e. most of the particles produced in LHC are measured at midrapidity \footnote{ The rapidity is defined as follows: $\eta=-\ln(\tan(\theta/2))$, where $ \theta$ is the angle between the particle three-momentum  and the positive direction of the beam axis. The $\eta \simeq 0$ is midrapidity range, $\eta \gg 1$  is forward, $\eta \ll -1$  is backward range. } range  $ | \eta|  <2.5$, while in the case of EAS	 most  of energy is carried by forward particles. To solve the problem  of extrapolation of  hadronic interaction properties to higher energies, we also  need  more LHC data  in the forward directions and for targets of intermediate mass to fill required phase-space for EAS. This is  proposed for example  in Ref.~ \cite{lhcoxygen} .

The Pierre Auger Observatory  is the world's largest operating detection system for the observation of UHECRs~\cite{auger} . The Observatory uses two techniques to measure properties of extensive air showers. The longitudinal development of air showers in the
atmosphere is observed with the Fluorescence Detector (FD) and the lateral distribution of particles in the
shower at ground level is recorded by the Surface Detector (SD). Charged particles and photons that reach the
ground are sampled with the SD array which consists of 1660 independent water-Cherenkov detectors (WCD)
equipped with photomultipliers to detect the Cherenkov light emitted in  water \cite{augerSD} . The SD array is spread
over an area of about 3000 km$^2$. The fluorescence light generated in the atmosphere by the charged particles
of the air shower through excitation of N$_2$ molecules is recorded by the FD consisting of 27 telescopes \cite{augerFD} .

In this paper  we overview the new Auger studies exploring the connection between the dynamics of the air shower and the multi-particle production, and how this knowledge can be translated into constraints of the high energy hadronic models as well as direct measurements, complementary to, and beyond the reach of, accelerator experiments.

\section{Air shower and its connections to hadronic interactions }	
Monte-Carlo simulations derived from the analytical study of electromagnetic showers can constitute a great tool to obtain a detailed description of the
evolution of particle cascades. However, a simpler approach through toy models, such as the one introduced by Walter Heitler in \cite{heitler} , is also capable of providing
an accurate prediction of some of the quantities that characterize electromagnetic (EM) showers without the need for high-performance computing.
%
 This  model, although  simplistic, reproduces
 two important features of air showers: the number of particles at shower maximum $N_{\mathrm{max}}$ is proportional to the energy of the primary particle, E$_0$  and 
 primaries with higher energy initiate showers that reach their maximum of development ($X_{\mathrm{max}}$) deeper in the atmosphere. The $X_{\mathrm{max}}$ is the  atmospheric depth where the longitudinal development of an air shower reaches the maximum number of particles. The development of an electromagnetic cascade is rather well understood. For example the electromagnetic 
 cascade takes more than $50$\%  of energy from 1$^{\mathrm{st}}$, 2$^{\mathrm{nd}}$ and 3$^{\mathrm{rd}}$  cascade generations, and   as  also  shown in Ref.~ \cite{em} 
the shapes of longitudinal profiles are well reproduced by the Gaisser-Hillas (GH) parametrization~\cite{gh} . In Ref.~ \cite{em} it was also reported, that   values of the  fitted parameters of the GH  parametrization to the data  from Pierre Auger Observatory are compatible (at the $2\sigma$ level) with parameters obtained  for all  hadronic interaction models~\footnote{For example,  in case of the so-called parameter $R$, which is sensitive to the injection of high energy $\pi^{0}$ in the start up of the shower.}.

Following up the  model for electromagnetic showers developed by
Heitler, Matthews worked out a similar approach by generalizing such model
to showers initiated by hadrons and nuclei \cite{hm}: a primary hadron interacts
with the medium and produces charged  and neutral
pions. The neutral pions rapidly decay to photons while charged pions travel
through a splitting length $d = X_{i} \ln(2)$, where $X_i$ is the interaction length of
charged pions ($X_i=120$ g cm$^{-2}$ in the air for pions) -- after  that they interact again to
produce a new generation of charged and neutral pions. At each step, one
third of the energy of the new generation is transferred to the electromagnetic
component through the neutral pion decay. Such a process carries on until the
energy per pion is smaller than the critical energy ($\xi_c =20$ GeV)  for which the decay length becomes smaller than $X_i$. This critical energy decreases
with increasing primary energy. 
At this point, charged pions  decay  to muons such that the number of muons $N_{\mu}$ is equal to the total number of charged pions at the last splitting length.  
The Heitler-Matthews model of hadronic air showers predicts the  following number of produced  muons: $N_{\mu}=A(E_0/(A\xi_c))^{\beta}$, where $A$ is the  primary mass, $E_0$ energy of cosmic ray and $\beta \simeq 0.9$. In fact, the parameter $\beta$ depends on the muon production.
  Detailed simulations of $\beta$ show further dependencies on hadronic-interaction properties, like the multiplicity, the charge ratio and the baryon anti-baryon pair production~\cite{ulrich} .   Simulations  and measurements confirmed that the produced number of muons, $N_\mu$, rises almost linearly with the primary energy $E_0$, and increases with a small power of the cosmic-ray mass $A$. Thus, the number of muons in an air shower is another powerful tracer of the mass and  is directly linked to the description of pion interactions in the EAS. Measurements of this  observable can effectively constrain the parameters governing hadronic  interactions and improve the accuracy of hadronic models.

As we already mentioned, the interpretation of EAS measurements heavily rely on the results obtained from Monte-Carlo simulations and consequently, on the features of the
hadronic models that are used to characterize particle interactions in simulation algorithms that describe the cascade development, such as CORSIKA
(COsmic Ray SImulations for KAscade) \cite{corsica} . In fact, different hadronic models may lead to important differences in expectations regarding some of the
EAS characteristics such as the number of muons or the depth of maximum development. 

Some of the hadronic model features that may affect Monte-Carlo simulations include the multiplicity, i.e. the number of secondary particles produced
at each hadronic interaction; the inelastic cross section, i.e the probability for an hadronic interaction to occur; or the inelasticity, i.e. the fraction of energy
carried by the parent particle after an interaction. Although these parameters tend to increase with the energy in p-air collisions, their absolute values differ
from one model to another as highlighted in Ref.~\cite{pierog} . The observed discrepancy between models at the highest energies is essentially
due to the lack of data in  the energy range of accelerator experiments. Three models, updated to take into account LHC data at 7 TeV, are usually taken into consideration: QGSJETII-04~\cite{qgsjetA,qgsjetB}, EPOS-LHC~\cite{epos} and SIBYLL2.3C~\cite{sibyll} . While the structure as well as the details of these models are beyond the scope of this paper, differences regarding the evolution of hadronic
interaction parameters from one model to another may be briefly discussed.

First of all, one noticeable feature is the fact that the extrapolations of the p-p cross section (measured up to  LHC energy)  at the highest energies in p-air collisions are consistent between
models while they diverge in the case of $\pi$-air collisions.  On the other hand, large variations are observed in the multiplicity for both types of interactions.
Similar behavior is seen in the evolution of the inelasticity. The divergence of these three parameters has serious consequences for the simulations performed
to reconstruct the features of observed EAS’s. The impact of the inelastic cross section on the development of EAS’s is important,  as it affects the depth
of first interaction $X_0$ and the number of particles (electrons and muons) at the surface, as well as their shower-to-shower fluctuations. Lower estimations may
significantly reduce the number of interactions occurring and therefore increase $X_{\mathrm{max}}$ (deeper cascades). Similarly, higher multiplicity increases the number of
particles produced at each interaction and the number of particles at ground level, and also has an influence on the $X_{\mathrm{max}}$. The effect of the extrapolation
of these parameters to the highest energies on some of the EAS observables such as the mean $X_{\mathrm{max}}$, the number of electrons and muons, as well as their
fluctuations, is well described in Ref.~\cite{ulrich} .

While the expected mean $X_{\mathrm{max}}$ suffers from uncertainties that are predominantly due to the nature of the first interaction, observables related to
muons, such as muon production depth and the number of muons at ground level, are directly linked to the description of pion interactions in
the EAS. Thus, measurements of these observables could effectively constrain the parameters governing these interactions and improve the accuracy of hadronic
models.

\section{Auger results }	
\begin{figure}[ht]
\centerline{\includegraphics[width=9.8cm]{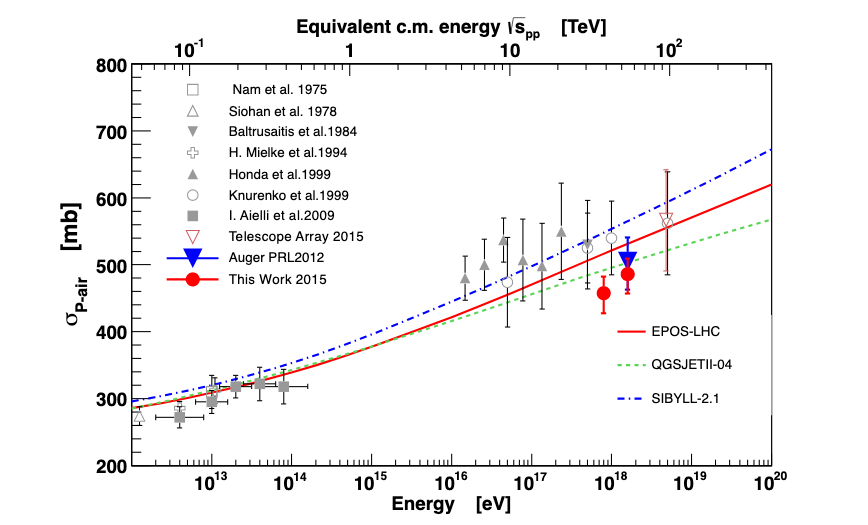}}
\caption{The  $\sigma_{p-\mathrm{air}}$ compared  to  other  measurements (see \cite{cross-sectionAuger1}  for references) and other model predictions. Figure taken from Ref.~\cite{cross-sectionAuger2} .
 \label{f11}}
\end{figure} 
 In this section we overview the  Auger studies to explore the connection between the dynamics of the air shower and the multi-particle production, and how this knowledge can be translated into constraints of the high energy hadronic models as well as direct measurements.
 
 \subsection{The p-air cross section }
 As  was shown for the first time  in the pioneering work of the Fly’s Eye Collaboration~\cite{cs-method}, the shape of the distribution of the largest values of the depth of shower  maximum, $X_{\mathrm {max}}$ is  sensitive  to  the  p-air  cross-section $\sigma_{p-\mathrm{air}}$.
  The $dN/dX_{\mathrm {max}}$ tail  falls off according to $ \exp(-X_{\mathrm max}/{\Lambda_{\eta}})$ with $\Lambda_{\eta}^{-1} \simeq \sigma_{\mathrm p-\mathrm{air}}$. In Ref.~\cite{cross-sectionAuger1} 
 the first measurement of the cross-section with high-quality events  seen by SD and FD (hybrid events) of the Pierre Auger Observatory  was  presented, while the extension of this analysis is in Ref.~\cite{cross-sectionAuger2}, where several sources of systematic uncertainty are revisited and analysis was performed for  four times larger statistics. In Fig.~\ref{f11}  updated measurements of the p-air cross section with Auger hybrid data are shown. The p-air cross section is now estimated in the two energy intervals in lg(E/eV) from 17.8 to 18 and from 18 to 18.5.  These energies are chosen so that they maximise the available event statistics, and at the same time lie in the region most compatible with a significant primary proton fraction.  The conversion from p-air to p-p by  using Glauber theory to get inelastic p-p cross-section was also reported in Ref.~\cite{cross-sectionAuger1} . The obtained inealistic p-p  cross-section agrees  quite well with an extrapolation from LHC energies to 57 TeV for a limited set of models~\cite{cross-sectionAuger1} .
 
\subsection{The muon  production depth }
\begin{figure}[h]
\centerline{\includegraphics[width=13.8cm]{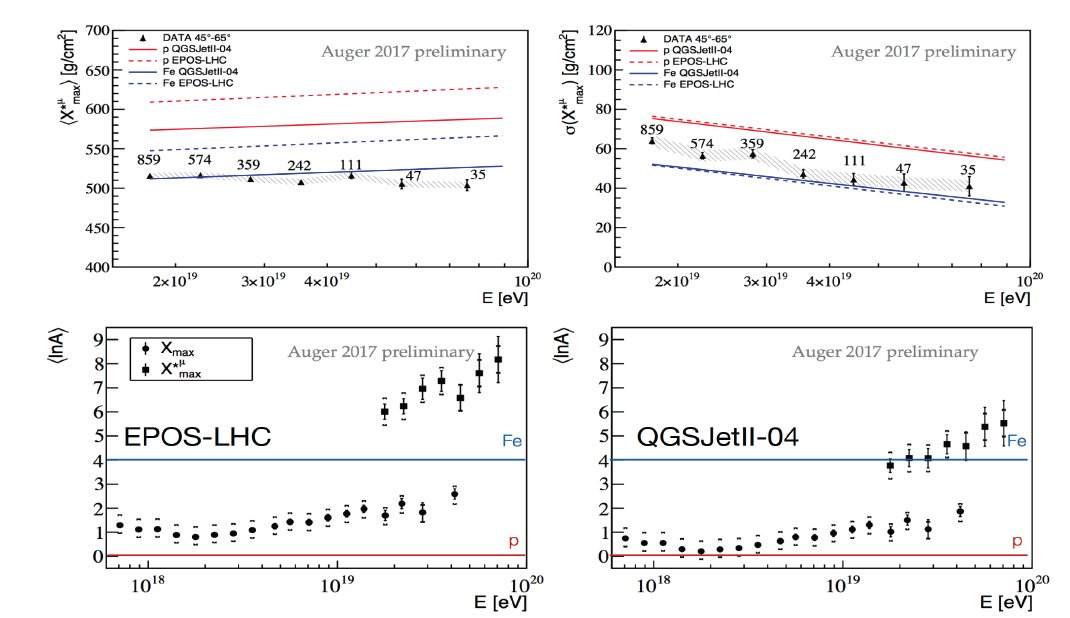}}
\caption{Upper panels:  $\langle X_\mu^{\mathrm {max}} \rangle$ (left) and the corresponding fluctuations (right) as a function of the primary energy. Data (black triangles) are shown with statistical (black line) and systematic uncertainties (gray band) and compared to simulations; Lower panels: The evolution with energy of $\langle  \ln A \rangle$ as obtained from the measured $ \langle X_\mu^{\mathrm{ max}} \rangle$ (squares). The results obtained for $\langle X_{\mathrm {max}} \rangle$ in Ref.~\cite{FDmax} (dots)  are also shown. EPOS-LHC (left) and QGSJetII-04 (right) are used as reference models. Square brackets correspond to the systematic uncertainties. Figures taken from Ref.~\cite{mpd2} .
 \label{f2}}
\end{figure}
The muons in EAS come mainly from  the decay of pions and kaons, and  most of them,  due to large momentum,  are produced along the shower axis, i.e. 
60\% of muons are produced within a radial distance from the shower axis smaller than 22 m, which is  much smaller  than the typical lateral distances of observation in experiments (few kms). 
Thus, we can   assume  quite well that muons travel in straight lines at the speed of light $c$ to the detector.
The arrival times of the muons allow the reconstruction of their geometrical production heights along the shower axis  and   to reconstruct a distribution of the  so-called  muon production depths (MPD), as  described in Ref.\cite{cal1,cal2}  and  later  in Ref. \cite{cal3} . The shape of the MPD distribution contains information about the evolution of the hadronic cascade, and  its maximum  is related  to the average energy of mesons  which reach the effective critical energy of the $\pi^{\pm}$/K mix.

Taking advantage of the large statistical sample provided by the SD array of Auger,  and  analysis of the FADC signals obtained with the WCD,
we can study the longitudinal development of the muonic component of EAS, as is shown in Ref.~\cite{mpd1,mpd2} .  The events with energy above  $10^{19.2}$ eV
and  with the  zenith angle between $55^{\circ}$ and $65^{\circ}$  were analysed,  and only  muons  recorded at more than 1700 m from the shower core were included in analysis. With the aim of obtaining useful physics information from the MPD distribution, for each event a fit of the muon longitudinal development profile with the Universal Shower Profile (USP)  function~\cite{gh} has been performed.  Of the four fitting parameters, $X_\mu^{\mathrm{max}}$ accounts for the point along the shower axis where the production of muons reaches its maximum as the shower develops through the atmosphere.
Later, according to the method presented in Ref.~\cite{conversion} ,  these $\langle X_\mu^{\mathrm{max}} \rangle$ measurements were converted  to  the mean logarithmic mass\footnote{The conversion  from $X_{\mathrm {max}}$ measuremed  to mean logarithmic mass   is  model dependent. }, $ \langle \ln(A) \rangle$ of the cosmic rays.

Figure~\ref{f2} (upper panels) displays the $\langle X_\mu^{\mathrm max} \rangle$, showing values compatible with QGSJEtII.04, and an extremely deep predicted value of $\langle X_\mu^{\mathrm max} \rangle$ for EPOS-LHC, making it lie  out of the p-Fe range. Figure ~\ref{f2} (lower panel) shows the calculated  mean logarithmic mass $\ln(A)$ from  $X_\mu^{\mathrm {max}}$, together with  mean logarithmic mass obtained from  $ \langle X_{\mathrm max} \rangle$ from FD measurements~\cite{FDmax}. 
In principle, for the  fixed interaction model, after conversion  to  $\ln(A)$ phase space, both type of data  should be   inside the p-Fe band.  However, it is  clearly seen that   $\ln(A)$ points from  $X_\mu^{\mathrm max}$ measurement  are outside p-Fe band. This   result is  a sign of a   problem in  the description of muon production in hadronic interaction models. As is  shown in Ref. ~\cite{mpd-InterA,mpd-InterB} , to  get consistent results between the mean logarithmic mass  $ \langle \ln(A) \rangle$ extracted from $\langle X_\mu^{\mathrm max} \rangle$ and  the one deduced from $ \langle X_{\mathrm max} \rangle$,  the following  changes are sufficient:  (1)  change of  the  baryon production:  baryons have smaller critical energy than muons, so  they reach deeper and do not produce muons; (2) taking into account  $\pi$-air diffraction processes: this slows down multiplicative process; or (3) consider changes in K/$\pi^{\pm}$  energy spectrum: bulk of mesons closer to critical energy.

\subsection{Muon studies with hybrid events }
\begin{figure}[ht]
\centerline{\includegraphics[width=8.8cm]{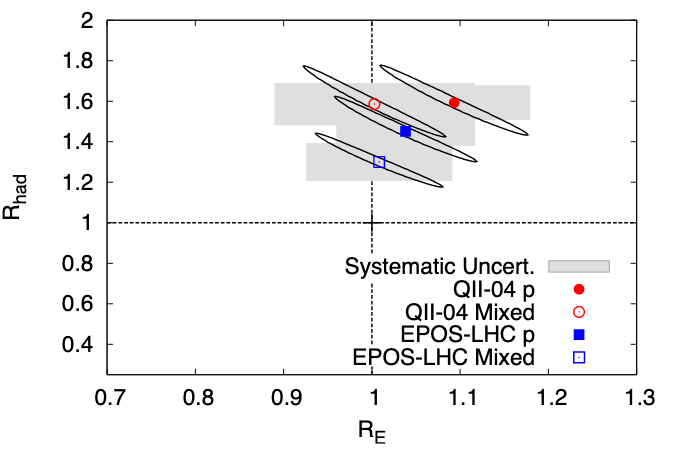}}
\caption{Phase space of the electromagnetic ($R_E$) and hadronic scaling factors ($R_{\mathrm had}$) for different
hadronic models and compositions. The 1$\sigma$ statistical and systematical uncertainties are
represented by the ellipses and grey boxes, respectively. Figure taken from Ref.~\cite{td} .
 \label{f33}}
\end{figure}

Another important topic is the testing of interaction models using dedicated simulations to
reproduce properties of real showers recorded at the Pierre Auger Observatory. These so-called top-down (TD) simulations have been performed 
for a set of high-quality hybrid events selected from the Auger data. The observed features of the events have been reproduced as closely as possible,
using different models and assuming various primary particles. The TD method finds a simulated shower, which has a particle distribution of its electromagnetic component along the shower axis (longitudinal profile) similar to the longitudinal profile of the real data shower (a reference profile). The reference longitudinal profile is directly linked with the electromagnetic component of the shower, so the considered method relies on the fact that this component is accurately simulated (is very well understood). As an output, the TD method provides a reconstructed event, in which the signals in detectors are determined using Monte Carlo simulations. The SD signals include also the muon contributions, which are the tracers of properties of the hadronic interactions. Comparison of simulated SD signals with the corresponding signals in the real data shower provides an opportunity to investigate the correctness of lateral distributions of the simulated showers. Since these distributions are sensitive to the hadronic interaction models, their analysis allows for indirect investigation of the interaction models at energies above those, for which the accelerator data are available. Such analysis was presented  by Auger~\cite{td} , using various hadronic models to compare with data obtained in the energy range 6 -- 16 EeV with zenith angles smaller than $60^{\circ}$.  Introducing a rescaling factor  $R_{E}$ and $R_{\mathrm had}$ for the measured electromagnetic and hadronic signals, respectively, the signals from Monte-Carlo simulations were matched to the recorded ones. Fig.~\ref{f33} shows the phase space of these rescaling factors for different hadronic models and different compositions (pure proton or mixed). While the electromagnetic signal does not need any rescaling for any scenario, the hadronic signal must be rescaled by a factor 1.3 -- 1.6 depending on the model considered. This means  that  the  observed  hadronic  signal in these UHECR air showers is significantly larger than predicted by models tuned to fit accelerator data, and the difference is due to a muon deficit in Monte-Carlo simulations  (i.e. "muon excess" in data). None of the models used in the simulations can reproduce the observed particle distributions. This indicates that our understanding of hadronic interactions at the highest energies is far from being satisfactory.

\subsection{Muon studies with inclined hybrid events}

\begin{figure}[ht]
\centerline{\includegraphics[width=14.8cm]{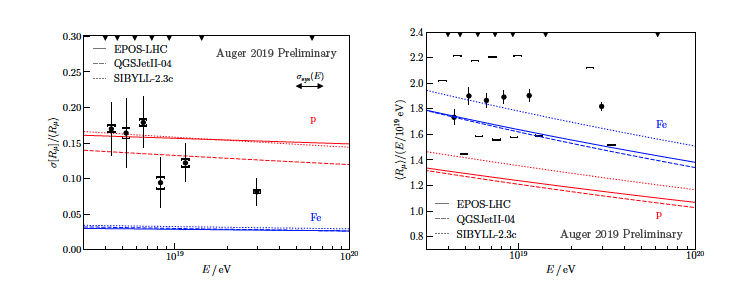}}
\caption{Shower-to-shower fluctuations (left) and the average number of muons (right) in inclined air
showers as a function of the primary energy. For the fluctuations, the statistical uncertainty (error bars) is
dominant, while for $\langle R_\mu \rangle$  the systematic uncertainty (square brackets) is dominant. The shift in the markers
for the systematic uncertainty in the average number of muons represents the uncertainty in the energy scale. Figure taken from Ref.~\cite{muoninclined2} .
 \label{f3}}
\end{figure}

A similar conclusion  about a muon excess in Auger data was reached in Ref.~\cite{muoninclined,muoninclined2} , where agreements between simulations
and data regarding the  muonic components were studied for  ultra-high energy cosmic rays with zenith angles between $62^{\circ}$ and $80^{\circ}$. For these inclined air showers, the reconstruction of the number of muons relies on the fact that the electromagnetic cascade is mostly absorbed in the atmosphere so that signals at the ground
are dominated by muons. An event selection is applied to ensure a high quality of the reconstruction. For the FD this means that only events measured during good atmospheric conditions are selected, but  for the SD, only events with reconstructed energies above 4 EeV and zenith angles above 62 EeV are accepted to ensure a triggering probability of 100\%  and to avoid contamination of the signals at the ground by the electromagnetic component. By fitting the normalization factor of a reference model for the muon density at the ground to the observed distribution of signals in the SD array, the number of muons can be extracted \cite{muoninclined}. In fact, the reconstructed quantity $\langle R_\mu \rangle$ is the total number of muons at the ground relative to the average of the total number of muons in a shower with primary energy $10^{19}$ eV. The reconstruction of the energy $E$ of the air showers is done by integrating the longitudinal shower profiles observed with the FD. The average number of muons  in inclined air showers as a function of the primary energy is shown in Fig.~\ref{f3} (right).  A notable difference between measurement and model prediction is clearly seen. 
Studies of inclined events give also the possibility to determine the shower--to--shower fluctuation in the number of muons $\sigma/\langle R_{\mu} \rangle$ by fitting a statistical model to the measured pairs of energy $E$ and number of muons $R_\mu$. However, as is shown in  Fig.~\ref{f3} (left) they agree quite well with  the model predictions. Since the average  number of muons in a shower and its  fluctuations depend on different stages in the shower development, this means that muon execes is caused rather by  the  later stages of the shower development than by the first hadronic interactions.

\begin{figure}[hb]
\centerline{\includegraphics[width=8cm]{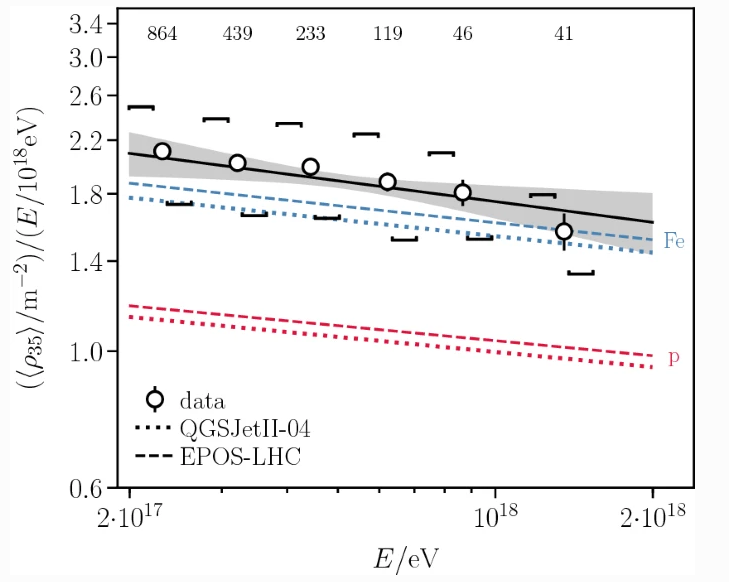}}
\caption{Energy-normalized muon densities $\rho_{35} /(E/10^{18}$ eV) as a function of E compared to expectations from simulations using EPOS-LHC (dashed) and QGSJetII-04 (dotted). Error bars denote the statistical uncertainties, while systematic uncertainties are indicated by square brackets. Figure taken from Ref.~\cite{umd} .
 \label{f1}}
\end{figure}

Another important result come from analysis of data from  the prototype array  of the so-called  Underground Muon Detector (UMD) upgrade of the currently working AMIGA detector. This detector will provide important direct measurements of the shower muon content and its time structure, while serving as verification and for fine-tuning of the methods used to extract muon information from the  WCD measurements. The performance and characteristics of the underground muon detectors~\cite{umd}, now being deployed, match these requirements. It is worth  mentioning that in a recent analysis using the data collected by the UMD engineering array, it was determined, for the first time with direct measurements of the muons, that the muonic densities in data are also  larger than those expected from simulations of showers between $10^{17.5}$ eV and $10^{18}$ eV~\cite{umd}, even after the update of the simulations to the latest hadronic interaction models, which include the data from the LHC (Fig.~\ref{f1}).

\section{Conclusions and summary }
The muon excess seen  by the Pierre Auger Collaboration is/was also seen in  several other experiments like
 HiRes/MIA~\cite{hires} , NEVOD-DECOR\cite{nevod}, SUGAR array~\cite{sugar} , Telescope Array~\cite{ta}. However,   experiments like KASCADE-Grande~\cite{cascade} and EAS-MSU~\cite{EASMU} reported  no discrepancy in the muon number around $10^{17}$ eV. In Ref.~\cite{hans} ,  after cross-calibration of the energy scales,
the observed muon densities were scaled by using the so-called $z-$scale and compared to expectations from different hadronic models, also for  data from IceCube~\cite{icecube} and AMIGA~\cite{amiga} .  While such densities were found to be consistent with simulations up to $10^{16}$ eV, at higher energies the muon deficit increases in several experiments~\cite{hans} . The muon problem  currently is one of the hot topics in  cosmic ray  physics, and  for a few years  some  attempts have been made to solve it, but up to now it has not been explained fully.  As we mentioned above,  it is because of   the inaccessibility of certain phase space regions to accelerator experiments, which are important for  the typical energies of EAS. On the other hand, even in the simple Mathews-Heitler model,  increasing the hadronic energy  fraction of interactions by about   5\%  per generation,   can  lead to about  30\% change in the  number of muons  after 6 cascade generations.  The formation of a Strange Fireball~\cite{sf} , String Percolation~\cite{sp} , Chiral Symmetry Restoration~\cite{csr} ,  increasing the inelastic cross section~\cite{cs} , or for instance resorting to Lorentz Invariance Violation~\cite{liv} could  also explain the muon excess seen in EAS. 

Ultra-high energy cosmic rays present a great opportunity to explore particle physics beyond the reach of accelerators. Our understanding of hadronic physics in the forward region and at the highest energies is a mere extrapolation to the unknown, and therefore is subject to uncertainties.
The mass interpretation of UHECR, on the other hand, needs EAS simulations which make use of the high energy hadronic models.

\section*{Acknowledgments}
The successful installation, commissioning, and operation of the Pierre Auger Observatory would not have been possible without the strong commitment and effort from the technical and administrative staff in Malargue, and the financial support from a number of funding agencies in the participating countries, listed at https://www.auger.org/index.php/about-us/funding-agencies. In particular we want to acknowledge support in Poland from National Science Centre grant No. 2016/23/B/ST9/01635 and  grant No.  2020/39/B/ST9/01398 and from the Ministry of Science and HigherEducation grant No. DIR/WK/2018/11.


\end{document}